\documentclass{article}
\usepackage{graphicx} 
\usepackage{authblk}
\usepackage{amsmath}
\usepackage{multirow}
\usepackage[sort, comma, numbers]{natbib}
\usepackage{pdflscape}
\usepackage[skip=5pt plus1pt, indent=40pt]{parskip}

\usepackage{geometry}
\usepackage{fancyhdr}
\title{Chirp Group Delay based Onset Detection in Instruments with Fast Attack}
\author[1]{S. Johanan Joysingh}
\author[1]{P. Vijayalakshmi}
\author[2]{T. Nagarajan}
\affil[1]{Department of ECE, Sri Sivasubramaniya Nadar College of Engineering, Chennai, India.}
\affil[2]{Department of CSE, Shiv Nadar University Chennai, India.}
\date{}

\begin{document}
\maketitle

\section{Abstract}
The onset of a musical note is the earliest time at which 
a note can be reliably detected. 
Detection of these musical onsets pose challenges 
in the presence of ornamentation such as vibrato, bending, 
and if the attack of the note transient is slower. 
The legacy systems such as spectral difference or flux and 
complex domain functions suffer from the addition of false 
positives due to ornamentation posing as viable onsets. 
We propose that this can be solved by appropriately improving
the resolution of the onset strength signal (OSS) and smoothening 
it to increase true positives and decrease false positives, respectively.
An appropriate peak picking algorithm that works well 
in unison with the OSS generated is also desired. 
Since onset detection is a low-level process upon which many other tasks 
are built, computational complexity must also be reduced. 
We propose an onset detection alogrithm that is a combination of
short-time spectral average-based OSS estimation, 
chirp group delay-based smoothening, and
valley-peak distance-based peak picking.
This algorithm performs on par with the state-of-the-art, 
superflux and convolutional neural networks-based
onset detection, 
with an average $\text{F}_{1}$ score of 0.88, 
across three datasets.
Subsets from the IDMT-SMT-Guitar, Guitarset, and 
Musicnet datasets that fit the scope of the work, 
are used for evaluation.
It is also found that the proposed algorithm is
computationally 300\% more efficient than superflux. 
The positive effects of smoothening an OSS, 
in determining the onset locations, is established by 
refining the OSS produced by legacy algorithms, 
where consistent improvement in onset detection 
performance is observed.
To provide insights into the performance of the proposed algorithms
when different ornamentation styles are present in the recording,
three levels of results are computed, 
by selecting different subsets of the IDMT dataset.

\vspace{0.25cm}
\begin{center}
\textbf{Keywords}: \textit{onset detection, chirp group delay, guitar, piano}
\end{center}

\section{Introduction }
\label{introduction}
An onset is a single instant chosen to mark the start of the transient  of a note, 
or the earliest time at which the transient can be reliably detected \cite{bello2005tutorial}. 
It can also be defined as the beginning of the attack phase
in the ADSR (Attack, Decay, Sustain, and Release) envelope of a note \cite{rodet2001detection}.
The process of detecting the onsets in a given piece of music is called onset detection.
Just like transients and impulses, onsets of most instruments 
(which include plucked and struck string instruments 
and pecussive instruments) are characterized by 
high energy across the frequency range. 
This is evident in a short-time Fourier transform.

Onset detection serves as a low-level process that is 
used across multiple varied higher level applications
such as, beat tracking \cite{mcfee2014better}, auditory
segmentation \cite{hu2007auditory}, 
and audio content analysis \cite{maka2016audio}. 
Low-level tools are expected to have higher accuracy 
and reduced computational complexity as their
performance will directly influence the performance 
of the higher level algorithms that depend on it.

In the real world, musicians often stylize a piece of 
music they are playing, 
and notes often may exhibit many distinct characteristics. 
In the guitar for example, popular articulations include
\begin{itemize}
  \item bending - varying the frequency of a note up or down by bending the string(s)
  \item vibrato - varying the frequency of a note up and down mildly and periodically
  \item slide - increasing or decreasing the frequency produced by 
    sliding the finger across the fret of the guitar horizontally, 
    and sliding on to the frequency of another note.
\end{itemize}
All these variations are evident in a short-time spectrum 
and can appear as viable onsets in the onset strength signal (OSS).
It is especially difficult for onset detection algorithms 
that depend on the variation in the short-time spectrum, 
since the difference between adjacent time frames will, 
in this case, yield considerable magnitude differences 
due to minor frequency changes within the duration of a note. 
These variations introduce false positives in the 
estimation of onsets, which must be reduced.

\subsection{Onset Detection Approaches}
\label{intro:literature}
Onset detection process can be summarized into a two step process. 
First, a two dimensional time-frequency representation is processed 
to produce a one dimensional signal along time, with a reduced sampling rate,
called the Onset Strength Signal (OSS) 
(also called the activation function or novelty function or
even onset detection function). 
In probabilistic terms, the amplitude of the OSS is directly proportional 
to the likelihood of an onset at a particular location. 
Peak picking is then performed on the OSS, to provide
the actual location of the onsets over time.
When processing real world music, a recording may be
mixed with noise or sounds from other instruments playing nearby.
In such cases, noise removal or source separation methods may be required 
to pre-process the signal before the OSS can be derived.
In a few cases, the process of splitting the signal into different sub-bands 
is also addressed as a pre-processing step \cite{bello2005tutorial}.

A summary of the major approaches in the literature 
used to derive the OSS can be found in 
\cite{bello2005tutorial}, \cite{dixon2006onset}, and \cite{rosao2011trends}.
They include time-domain methods, spectral-domain methods,
pitch-based methods, probabilistic methods, 
and methods based on neural networks.

\subsubsection{Time-domain methods}
In time-domain-based OSS estimation, the energy or 
magnitude envelope of the signal is derived.
Percussive sounds are generally characterized by a high 
rise in energy at the beginning of the note, 
and this property is utilized by the time-domain methods. 
But not all instrument onsets show such characteristics 
in the time-domain, and hence its application is limited.

\subsubsection{Spectral-domain methods}
Spectral-domain methods work based on the energy, 
magnitude or average envelope extracted from a time-frequency
representation derived using the short-time Fourier Transform (STFT). 
The proposed work is categorized under this
approach, hence comparatively more details are provided below. 
Most popular among spectral methods are the spectral difference or flux, 
phase deviation, complex domain, and superflux.

The intuition behind the spectral flux is that 
the amount of change or variation in the magnitude spectrum,
which is high in and around the onset locations,
can be identified by finding the difference between 
adjacent frames of a spectrogram.
Here, if L1 norm is computed, it is called spectral flux, and 
if L2 norm is computed, it is called spectral difference
\cite{rosao2011trends}. 
The differenced spectrogram is then summed across
frequency for each time frame to produce the OSS.
The phase deviation function \cite{bello2003phase}, 
makes use of the fact that frequency components 
of the new notes are unlikely to be in phase 
with the previous note. 
The complex domain method works based on 
the difference between the actual and the expected value 
of the spectral magnitude and phase, 
which is estimated by assuming constant magnitude and 
constant rate of phase change \cite{duxbury2003combined} \cite{duxbury2003complex}.
Temporal reassignment using group delay to improve the 
temporal resolution of the spectrogram can be found 
in \cite{o2018improved}.
These methods are prone to false positives in the results,
when there are some forms of ornamentation in the recording. 
The state-of-the-art in this category is superflux \cite{bock2013maximum},
which is proposed as a solution to handle vibrato. 
It works well, but uses a maximum filter to handle vibrato,
which increases the computational complexity.
In \cite{bock2013local}, the authors add to superflux by 
weighting it using the local group delay, 
which is the difference along the frequency axis of 
an unwrapped phase spectrogram.
This weighing reduces the impact of vibrato and tremolo.
Some of these methods employ sub-band filtering,
where an OSS is derived from each of the sub-bands
\cite{bello2005tutorial} \cite{dixon2006onset}.
Log-filtering is also performed in some cases to accentuate 
the set of fundamental frequencies that correspond to the notes
\cite{bello2005tutorial}. 


The distinct advantages of using spectral domain methods 
over the other types such as probabilistic methods and 
methods based on neural networks are that, 
(i) it does not require training data, and 
(ii) it offers a good trade-off between 
onset detection performance and computational complexity.

\subsubsection{Pitch-based methods}
For instruments with slow attack (pitched non-percussive), 
or in other words, instruments for which there is no
definite rise in amplitude in the attack phase, 
like a violin for example, energy- or magnitude-based methods
mentioned so far do not work well, and hence pitch-based 
approaches are preferred \cite{collins2005using}.
In such instruments, a sequence of notes is played by 
sliding from one note frequency to the other. 
Hence, the onset is detected based on the change in note frequency. 

\subsubsection{Probabilistic methods}
These methods rely on building a probabilistic model  
for the onsets and/or the steady state segment of the signal.
Few methods employ a strategy that looks for sudden change
in the characteristics of the signal with respect to a single model, 
while other methods look for the best likelihood with respect
to multiple models \cite{bello2005tutorial}.
In \cite{degara2011onset}, hidden Markov model is used to
model the rhythmic structure of the signal.

\subsubsection{Neural Networks-based methods}
Recent algorithms based on Neural Networks, 
take advantage of the vast processing power
available. 
In \cite{marolt2002neural}, the authors use a network 
of integrate-and-fire neurons along with a multi-layered 
perceptron exclusively for the task of peak-picking, 
but use spectral-domain methods to derive the OSS.
In \cite{lacoste2006supervised}, authors propose two algorithms
based on the use of single and multiple
feed-forward neural networks (FNN). 
An onset detector that uses recurrent neural networks (RNN), 
can be found in \cite{bock2012online}.
In \cite{eyben2010universal}, a universal onset detector
using bidirectional long short-term memory RNN is presented. 
An example of onset detection using Convolutional Neural Networks 
(CNN) can be found in \cite{schluter2014improved}.
This method is also considered the state-of-the-art, 
and hence used for comparison in our evaluation.

A shortcoming of the probabilistic and neural networks-based 
approaches is the fact that they either need to be 
trained on annotated datasets which may not always be available,
or they need pre-trained models, which may or 
may not suit the task at hand.
Developing large annotated datasets for music is 
not a trivial task \cite{su2015escaping}.

\subsection{Proposed Solution}
\label{intro:proposed}

Generally, onset detection is accomplished in two stages, namely,
\begin{enumerate}
    \item Obtaining the OSS function
    \item Picking the peaks to estimate the onset locations
\end{enumerate}

\begin{figure*}[h]
  \centerline{\includegraphics[width=0.5\textwidth]{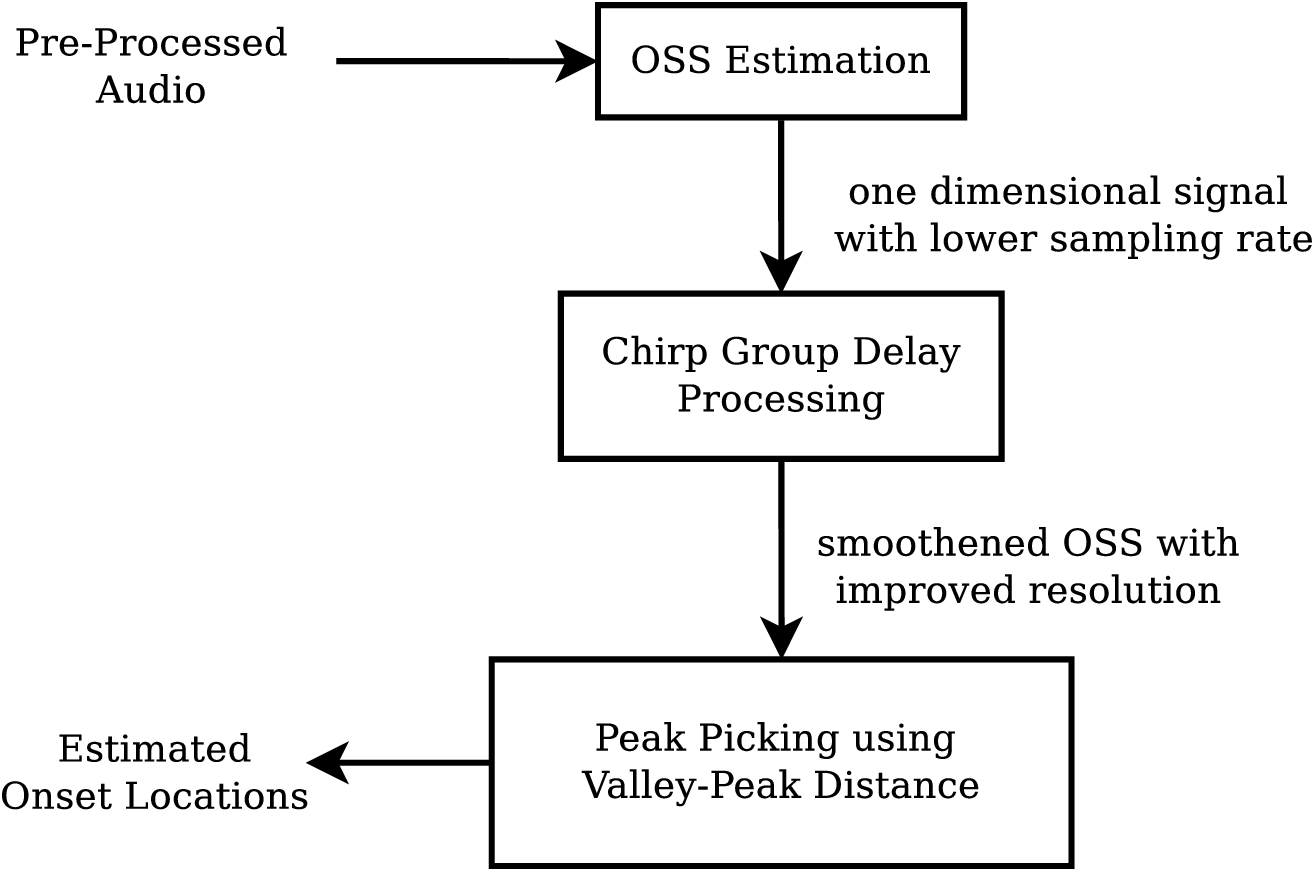}}
  \caption{Proposed Solution - Overview} 
  \label{fig:proposed_solution}
\end{figure*}

In the current work, we propose that the onsets can be
detected by estimating a short-time spectral average-based OSS, 
which is further processed by a 
chirp group delay-based smoothening algorithm,
whose peaks are picked by a 
valley-peak distance-based peak picking algorithm.
The chirp group delay-based smoothening algorithm
shines in achieving two things,
\begin{enumerate}
  \item smoothening the OSS to reduce the load 
  on the peak picking algorithm, 
  thus reducing the false positive rate
  \item providing better resolution to the 
  actual large peaks and valleys of the OSS, 
  thus increasing the true positive rate.
\end{enumerate}
Here, false positives are the frames 
falsely identified to contain onsets 
and true positives are the frames 
correctly identified to contain onsets.
Fig. \ref{fig:proposed_solution} is a schematic 
description of the proposed solution.

The scope of the current work is limited 
to recordings from 
instruments that have noticeable transients, 
whose magnitude envelope
either derived from the time-domain or 
the time-frequency representation,
can be used to detect the onsets.
All plucked string and struck string instruments, 
almost all types of keyboards, 
and some types of woodwind instruments will fit this category.
The datasets chosen in the current work 
address these categories broadly. 
More details can be found in Section \ref{corpus}.

\subsection{Organization of the Paper}
\label{intro:organization}
The characteristics of the OSS, and how it is processed in the legacy 
and state-of-the-art algorithms related to the current work 
are detailed in Section \ref{the_oss}.
Details of the proposed algorithms 
are explained in Section \ref{proposed_algorithms}. 
The theory and intuition behind using chirp group delay, 
along with its apparent strengths are detailed in 
Section \ref{chirp_group_delay_algorithm}. 
The proposed peak picking algorithm is explained in 
\ref{peak_picking}.
Finally in Section \ref{evaluation}, the dataset, scoring 
methodology, experimental setup and 
the experimental results are detailed.


\section{The Onset Strength Signal}
\label{the_oss}
An ideal onset strength signal (OSS) is a function of time in which 
the peaks coincide with the actual onsets in the signal. 
It is usually derived from a time-frequency representation 
of the signal $X(n,k)$ and hence has a lower sampling rate --- 
the number of samples being equal to the number of frames 
obtained from the signal in the time-frequency representation. 
Here $n$ and $k$ represent the number of time frames 
and the number of frequency bins respectively. 
Briefly explained below are onset strength signals
derived from various methods. 
They are used in the current work for comparison, 
hence more details are provided.

\subsection{Complex Domain Function}
The complex domain function (CDF) introduced in \cite{duxbury2003complex}
relies on measuring the joint departure 
of steady state behavior in both the amplitude and phase spectra.
The steady state amplitude and phase with respect to 
the current time frame, encompassed in $X_S(n,k)$,
is estimated based on the previous time frame, 
assuming a constant amplitude and constant rate of phase change. 
It can be given by, 
\begin{equation}
    X_{S}(n,k) = |X(n-1,k)|e^{\phi(n-1,k) + \phi'(n-1,k)},
\end{equation}
where $|X(n,k)|$ and $\phi(n,k)$ are the short-time magnitude and phase spectra, respectively.
The sum of the magnitude of the deviation of this steady state estimate 
from actual values, across all frequency bins, gives the magnitude of 
the onset strength signal of the complex domain function $\mathrm{CD}(n)$.
\begin{equation}
    \mathrm{CD}(n) = \sum_{k=0}^{\frac{N}{2}-1} |X(n,k) - X_S(n,k)|,
\end{equation}
where $N$ is the total number of frequency bins. 
In some cases the output is rectified using a 
half wave rectifier (HWR).
This function is named rectified complex domain function,
and is reported to yield performance better than the 
vanilla complex domain function \cite{dixon2006onset}. 
However, rectified complex domain function is not used in the
current work since it did not 
offer any improvements to complex domain.
This maybe due to the presence of ornamentation
in the dataset.

\subsection{Spectral Flux Function}
Spectral flux \cite{masri1996computer} is by far the most popular algorithm
for detecting onsets. 
It yielded the best performance in literature until it was superseded 
by its variant known as superflux. 
Mathematically, it is the sum of the positive differences between 
corresponding frequency bins of adjacent frames 
in a short-time magnitude spectrum. 
Negative differences are ignored using a half wave rectifier. 
The spectral flux function $\mathrm{SF}(n)$ is given by, 
\begin{equation}
    \mathrm{SF}(n) = \sum_{k=0}^{\frac{N}{2}-1} \mathrm{H}(|X(n,k)| - X(n-1,k)),
\end{equation}
where $\mathrm{H}$ is the half wave rectifier function.
It can also be visualized as a distance between successive 
short-term Fourier spectra, treating them as points in 
an N-dimensional space \cite{bello2005tutorial}.

\subsection{Superflux Function}
One of the state-of-the-art techniques in onset detection
is the Superflux algorithm. 
It is the spectral difference of the 
maximum filtered short-time Fourier spectra $X_{max}(n,k)$. 
It was introduced in \cite{bock2013maximum}, and aims to suppress
the false positives caused by the presence of vibrato in music. 
It adds a few components to the existing spectral flux algorithm. 
Log-filtering is performed with 24 filters per octave from 
30Hz to 17kHz to filter and retain frequency components 
centered around musical pitches.
The maximum filtering operation performed across frequency bins
to stabilize vibrato is given by
\begin{equation}
    X_{max}(n,k) = max(X(n,k-1:k+1)).
\end{equation}
This time-frequency representation is then used to calculate 
the spectral difference to produce the 
superflux function $\mathrm{SF^*}(n)$. 
\begin{equation}
    \mathrm{SF^*}(n) = \sum_{k=0}^{\frac{N}{2}-1} H(X_{max}(n,k) - X_{max}(n-\mu,k)),
\end{equation}
where $\mu$ is an offset calculated based on 
window size, window shape, and hop size.
At this stage, the differences, between time frames,
that are usually caused by vibrato are 
nullified to a greater extent due to the maximum filtering.
Half wave rectification is also performed.

\subsection{CNN-based OSS Estimation}
\label{convolutional}
Another state-of-the-art in onset detection is using 
convolutional neural networks as explained in
\cite{schluter2014improved}.
In this method, three spectrograms are computed, 
each with a different window size, but the same frame rate.
Using the same frame rate ensures the number of frames 
across these three spectrograms are the same.
Different window sizes would mean different 
frequency resolutions, but they are reduced to the
same number of frequency bins by using the
same log-filter across the three spectrograms.
Convolutional neural network is trained by assuming that
the three spectrograms are three channels of the input 
(as in computer vision applications, involving color images).
This trained model is then used to generate the OSS.
The pre-trained model provided by authors of
\cite{schluter2014improved} as part of the 
madmom python library \cite{bock2016madmom}
is used in the current work for evaluation purposes.

\subsection{OSS - The Challenge}
A common issue that is faced in using all the OSS algorithms is 
the addition of spurious onsets, generally called false positives.
Usually in existing methods, 
the threshold parameters in the peak picking algorithm 
are varied to reduce spurious onsets. 
But this can introduce false negatives. 
False negatives are time frames in which an onset is 
present but is not detected.
The current work aims to resolve this issue.


\section{Proposed Algorithms}
\label{proposed_algorithms}

To address the challenges with respect to the estimation 
of the OSS and the detection of onset locations,
as described in the previous sections, 
a chirp group delay-based onset detection algorithm 
is proposed. 
Here, the OSS is estimated by 
short-time spectral average (STSA), 
and the estimated OSS is processed by 
chirp group delay-based smoothening algorithm (CGD), 
and the peaks are picked by the valley-peak distance-based
peak picking algorithm (VPD). 
In the following sections, these three proposed elements
are elaborated individually.
The contribution of each element to the 
over all onset detection process is also explained.

\subsection{Short-Time Spectral Average-based OSS Estimation}
\label{shortimespectralaverage}
A simple method to estimate the OSS is by computing the sum 
or average of the magnitude spectrum for each time frame. 
In the current work we compute the short-time spectral average 
$\mathrm{SA}(n)$, to obtain the OSS. 
It is the average of the short-time Fourier spectrum 
across frequency bins at each time frame. 
It can be defined as
\begin{equation}
    \mathrm{SA}(n) = \frac{1}{N/2} \sum_{k=0}^{\frac{N}{2}-1} X(n,k).
\end{equation}

\begin{figure}[ht]
  \centerline{\includegraphics[width=0.65\textwidth]{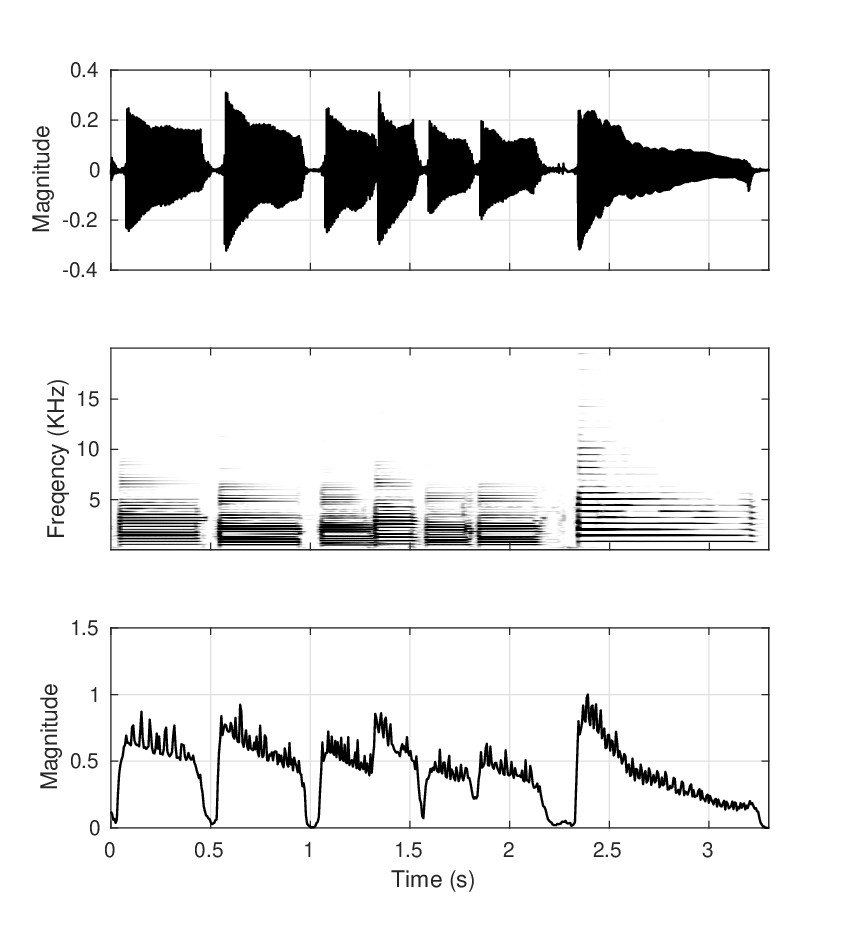}}
  \caption{Figure illustrates the, (a) time-domain signal, 
  (b) spectrogram of (a), and
  (c) the OSS computed using short-time spectral average, 
  of seven notes}
  \label{fig:onset_detection}
\end{figure}

The reason for this choice of an OSS function is as follows.
First, as discussed earlier in Section \ref{the_oss},
when there are ornaments in the recording, 
and when a spectral difference based algorithm is used,
spurious peaks are introduced in the OSS.
The process of differencing adjacent time frames of
a short-time spectrum creates these spurious peaks.
Using a spectral-average function, without spectral differencing,
provides one level of protection against spurious peaks in the OSS.
Secondly, onsets are characterized by energy across the spectrum 
and not just musical pitches. 
This contrasts it from the decay, sustain and release segments 
of the note.
Applying a log-filter on the spectrum,
as  is common in literature, 
removes part of this information
that is specific to the onsets. 
Hence it is avoided in the current work.
Thirdly, an OSS that is apt for the proposed 
chirp group delay-based smoothening algorithm and 
the proposed valley-peak distance-based peak picking
algorithm is desired. 
Generally, OSS are designed to have 
strong peaks in the location of the onset, since 
this makes peak-picking simple.
But, in the context of the other two proposed 
algorithms for smoothening and peak picking, 
an OSS that rises and falls along with 
the average spectral magnitude works well.
Lastly, it is computationally simpler.
Even in combination with the chirp group delay-based
smoothening algorithm, the computational complexity 
of estimating and processing the OSS is 
less than other state-of-the-art solutions.
This is discussed in detail in Section \ref{evaluation}.
The result of computing short-time spectral average 
over a small segment of music containing seven notes
is shown in Fig. \ref{fig:onset_detection}.

\subsection{Chirp Group Delay-based Smoothening Algorithm}
\label{chirp_group_delay_algorithm}

The group delay spectrum can be defined as the negative derivative of 
the Fourier transform phase spectrum.
It is known for its improved resolution,
due to its additive nature 
as compared to the multiplicative nature of the FT derived magnitude
spectrum \cite{murthy1991formant, murthy2011group, 
bozkurt2007chirp, vijayalakshmi2007acoustic, nagarajan2003segmentation}.
Group delay spectrum is said to have good resolution 
even in the presence of anti-resonances that tend to pull down 
resonance peaks in a magnitude spectrum, 
in addition to closely spaced resonances.
The adoption of group delay spectrum due to these 
advantages, and assumption of 
a time-domain signal as the magnitude spectrum,
by one or more of the authors of the current work,
can be found in
\cite{sripriya2015estimation}, \cite{rachel2015estimation}, 
\cite{rachel2017estimation}, \cite{nagarajan2004subband} and \cite{rachel2018significance}.
It is also used in \cite{kumar2015musical}, for onset detection
on percussive instruments.

The group delay spectrum computed on any circle 
with radius $r \neq 1$ (or in a spiral path in the z-plane), 
rather than on the unit-circle, may be considered 
a {\bf{chirp group delay spectrum}}. 
It was originally introduced, as a way to remove the sharp spikes 
introduced when computing group delay spectrum of a segment 
of speech signal, in \cite{bozkurt2007chirp}. 
These spikes in the spectrum are due to singularities on the 
unit-circle, and they are visible in the resultant spectrum 
since the measurement is made on the unit-circle. 
Instead, if the group delay spectrum is computed on a circle with radius 
greater than 1, these spikes will not be prominently visible. 
This, in a sense, can be thought of as 
smoothening the group delay spectrum.

In the current work, we utilize the resolving and 
smoothening properties of the chirp group delay spectrum 
and apply it to the OSS. 
Computing the spectrum outside the unit circle
ensures a smoothened OSS function, and 
considering the group delay spectrum, 
ensures better resolution. 
For this, we assume that the unprocessed OSS is 
the Fourier transform magnitude spectrum of an arbitrary signal 
and hence derive the chirp group delay spectrum for this 
assumed arbitrary signal, which would have the desired characteristics, 
namely, better resolution and smoothness.

Computing the Fourier transform at a radius greater than 1 is given by
\begin{equation}
\label{eq2}
  X(z)|_{z=re^{j\omega}} = X(r,\omega) = \sum_{n=-\infty}^{\infty} (x[n]r^{-n}) e^{-j \omega n}.
\end{equation}
As given in eq. (\ref{eq2}) , when $r$ is greater than 1, 
the signal is actually multiplied with $r^{-n}$, 
which corresponds to a signal that decays 
faster than the original signal. 
In the z-domain, it is equivalent to pushing the poles corresponding to the signal,
inside and towards the origin, which will result in a smoothened group delay spectrum.

\begin{itemize} 
  \item Consider a given OSS function $O(k)$ as half (from $0$ to $\pi$) of the discrete Fourier 
transform (DFT) of an arbitrary signal. 
  \item Symmetrize the OSS function to fully resemble DFT. Let us consider this symmetrized
OSS function as $X(k)$. 
  \item Compute the inverse DFT of the symmetrized signal. Since $X(k)$ is
a positive and symmetrical function, the inverse DFT of such a function will result in an even function, $x(n)$.
  \begin{equation}
      x(n) = \mathrm{IDFT}\ [X(k)]
  \end{equation}
  \item Consider the causal portion of $x(n)$. Let us denote it as $x_{c}(n)$. 
    \begin{equation}
      x_c(n) = x(n) \qquad n >0
    \end{equation}
  \item Compute the DFT of $x_{c}(n)$ at a radius greater than 1.
  \begin{equation}
  \begin{split}
      X_c(k, \omega) =\ & \mathrm{DFT}_{r>1}\ [x_c(n)] \\
      =\ & \sum_{n=0}^{N-1} [x_c(n) r^{-n}] e^{-j\omega n} \qquad r > 1 \\
  \end{split}
  \end{equation}
  \item Let the phase spectrum be $\theta_{xc}(k)$. 
  Compute the group delay spectrum from $\theta_{xc}(k)$, which will result in chirp group delay spectrum, $\tau_{xc}(k)$. 
  \begin{equation}
    \tau_{xc}(k) = -\frac{d}{dk} \theta_{xc}(k)
  \end{equation}
\end{itemize}

\begin{landscape}
\begin{figure}[]
  \vspace{1.2in}
  \includegraphics[width=1.50\textwidth]{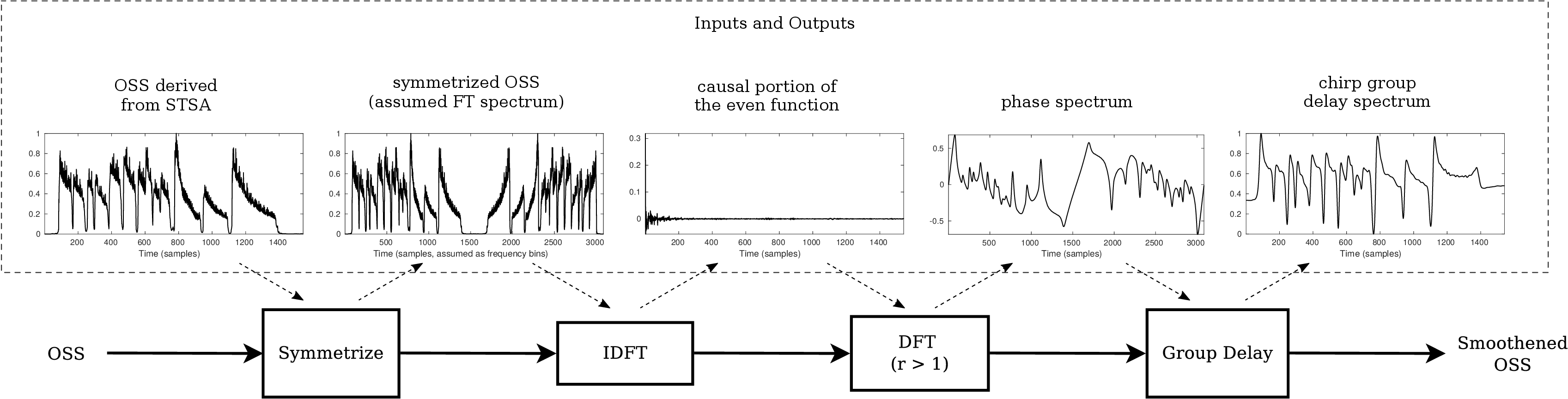}
  \caption{The chirp group delay algorithm for smoothening the OSS.
  The OSS shown here is derived using short-time spectral average.
  The inputs and outputs at each stage
  are also plotted above the blocks.}
  \label{fig:chirp_algorithm}
\end{figure}

\begin{figure}
  \vspace{0.2in}
  \centerline{\includegraphics[width=1.5\textwidth]{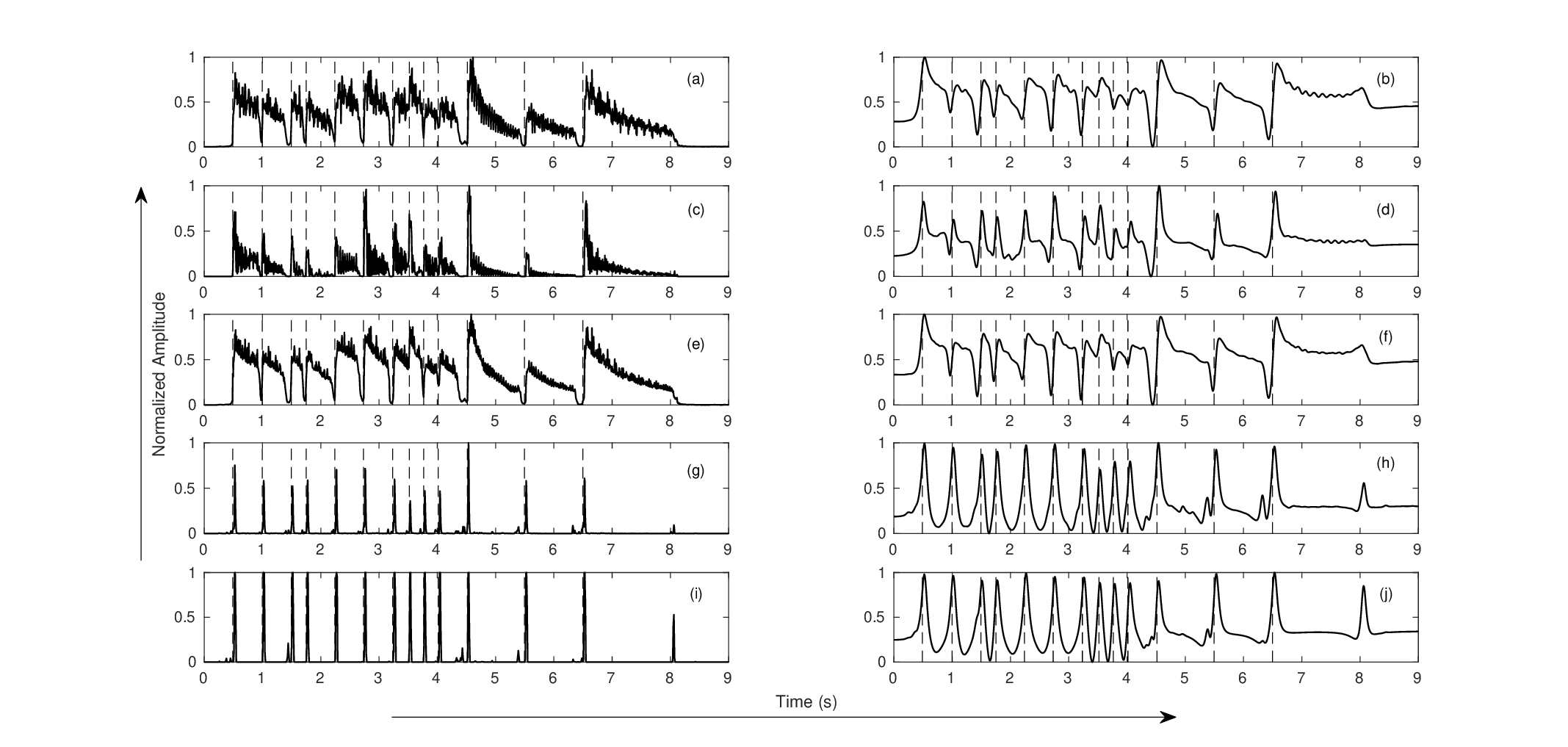}}
  \caption{Chirp group delay smoothening applied to OSS derived using the functions considered in the current work. The input is a lick from the IDMT dataset. Plots on the left, (a), (c), (e), (g), (i) show OSS derived using complex domain method, spectral flux, short-time spectral average, superflux and CNN-based OSS estimation respectively. Plots on the right, (b), (d), (f), (h), (j) show their corresponding OSS smoothened by Chirp group delay.}
  \label{fig:chirp}
\end{figure}
\end{landscape}

Fig. \ref{fig:chirp_algorithm} illustrates the 
chirp group delay algorithm for smoothening the OSS.
The result of computing the chirp group delay for the 
various OSS functions can be observed in Fig. \ref{fig:chirp}. 
It can be seen that the chirp group delay smoothens the OSS functions
and also improves the resolution of the valleys, 
resulting in a function that is more suitable for peak picking. 
The complex domain and the short-time spectral average functions
are not that sensitive to change in the time-frequency representation, 
when compared to the spectral flux and superflux functions. 
This can be observed in the OSS functions of the latter 
which contain sharp decays.
This sensitivity is good, but not always positive 
in the case of the spectral flux function, as this is the main reason 
for it being prone to detecting spurious onsets. 
From subplots (g), (h), (i) and (j) in Fig. \ref{fig:chirp}, 
it can be seen that the minor spurious peaks in the OSS produced 
by superflux and CNN-based OSS estimation,
are boosted by the smoothening process, 
hence being unable to contribute positively to the results.
Hence, the chirp group delay-based smoothening algorithm was not applied
to the OSS produced by superflux or CNN-based OSS estimation. 

\subsection{Valley-Peak Distance-based Peak Picking Algorithm}
\label{peak_picking}
The final peak-picking step, with its thresholds and constraints,
can have a large impact on the onset detection performance \cite{dixon2006onset}. 
An extensive study devoted to this 
can be found in \cite{rosao2012influence}.
In the current work, a valley-peak distance-based peak picking algorithm (VPD) is proposed. 
It decides the validity of a detected peak by measuring the 
distance between the peak and its preceding valley. 
Owing to the nature of the Chirp Group Delay algorithm,
which boosts this distance for better resolution, 
this idea works well in tandem. 
In Section \ref{evaluation}, it is compared implicitly with the peak picking algorithm 
used in \cite{dixon2006onset} and \cite{bock2013maximum}. 

The proposed peak picking algorithm makes a decision based on 
the relative distance of valleys and its corresponding peaks. 
The algorithm is as follows:
\begin{itemize}
  \item Find the locations of the peaks $p_x(m)$ and 
  valleys $v_x(m)$ in the signal $x(n)$.
  \begin{equation}
  \begin{split}
    p_x(m) = n &\quad \mathrm{if}\ x(n-1) < x(n) > x(n+1) \\
    v_x(m) = n &\quad \mathrm{if}\ x(n-1) > x(n) < x(n+1) \\
               &\quad \mathrm{for}\ n = 1,2,...,N-1,
  \end{split}
  \end{equation}
  where $N$ is the length of the signal.
  \item Calculate the valley-peak distances $d_{vp}(m)$ by 
  finding the difference between the amplitude of 
  each pair of peak and preceding valley.
  \begin{equation}
  \begin{split}
      d_{vp}(m) =& x(p_x(m)) - x(v_x(m)) \\
                & \mathrm{for}\ m = 1,2,...,M, 
  \end{split}
  \end{equation}
  where $M$ is the total number of peaks (or valleys) 
  detected in the signal.
  \item Calculate a threshold $T$ that is proportional to 
  the maximum valley-peak distance in the current signal.
  \begin{equation}
    T = \mu \operatorname*{max}_{m = 1,2,...,M}[d_{vp}(m)],
    \label{eq:threshold}
  \end{equation}
  where $\mu$ is a scaling factor chosen to maximize the performance. 
  A value of $0.75 < \mu < 1$ is appropriate.
  \item The onset locations $O$ are given by the valleys, 
  that precede the detected peaks, which have 
  valley-peak distance greater than the threshold.
  \begin{equation}
  \begin{split}
    O(l) = v_x(m) &\quad \mathrm{if}\ d_{vp}(m) > T \\
                  &\quad \mathrm{for}\ m = 1,2,...,M, 
  \end{split}
  \end{equation}
  where $l = 1,2,...,L$, and $L$ is the total number of onsets detected from $x(n)$.
\end{itemize}

The proposed peak picking method works particularly well in unison with 
the chirp group delay smoothened signals.
It also uses only one threshold parameter 
and is computationally less complex.
Smaller the number of parameters to tweak, 
smaller is the search space, when used with a new dataset.

\section{Evaluation}
\label{evaluation}

\subsection{Datasets}
\label{corpus}
The first dataset used in the current work is the IDMT-SMT-Guitar dataset \cite{kehling2014automatic}. 
This is the primary dataset, and was selected due to the following features: 
(i) well thought out annotations, 
(ii) availability of monophonic and polyphonic licks,
(iii) presence of various ornamentation such as vibrato, bending, slides, and dead notes 
(iv) use of multiple picking styles such as fingered, picked, and muted,
(v) use of multiple guitars, and 
(vi) use of clearly encoded filenames for categorization.
The authors of the dataset also intend it for onset detection.
It can also be stated that it has generality 
(containing a good representation of real world sounds) and 
quality (quality of the recorded audio and accuracy of the transcription) 
--- the required characteristics of a good dataset \cite{su2015escaping}. 

The dataset consists of four sub-datasets. 
Only data subset 2 is used for the current work. 
This is due to the following reasons:  
(i) dataset 1 consists of only a single note in each audio file
and onset detection would not be very relevant here, 
(ii) dataset 3 consists of very few musical pieces with no expression or styles, 
(iii) dataset 4 is mainly intended for genre identification, and 
(iv) dataset 2 has all the six features mentioned in the previous
paragraph, which is ideal for onset detection.
The contents of this data subset, from here on addressed as 
just the `IDMT dataset', can be broadly split into two
categories --- licks and scale-like pieces. 
A lick is a stock pattern or phrase consisting of a short series of
notes used in solos and melodic lines and accompaniment. 
These contain various expressions such as bending,
slide, vibrato, dead notes, lage, and harmonics in various combinations
(discussed later in Table \ref{tab:results_licks}).
Lage is a fingering variant whereby a different finger position is used 
on the fret-board of the guitar, to play the same notes. 
The scale-like pieces consist of pieces that sound like test runs 
for a particular guitar, yet they are simply called 'scales' 
in the current work for lack of a better name. 
These include chromatic scales, arpeggios (normal and muted) 
and various bending and sliding trials. 

The other two datasets used in the current work are the 
Guitarset \cite{xi2018guitarset} and Musicnet \cite{thickstun2017learning}. 
The former contains recordings from a steel-string acoustic guitar, 
while the latter consists of recordings of multiple instruments 
among which those of piano solo are used in the current work. 
The original scope of these two datasets is Automatic Music Transcription (AMT) 
and hence,  they contain a very large amount of data (considering the current task). 
A subset is formed by selecting a small portion from each of them.
Specifically, the subset formed from the Guitarset contains 763 onsets, 
while that from Musicnet contains 842 onsets.
This size is chosen based on existing similar work on onset detection.
For comparison, in the tutorial paper \cite{bello2005tutorial}, 
a total of 1065 onsets are used, 
being split into multiple categories based on their nature.
Upon random checks of the annotation in the dataset, 
it was noticed that the labels in these two datasets 
contained some minor errors.
Reference labels must have both good accuracy and consistency.
Since onset annotations are almost always done by-hand, 
they are reviewed and corrected manually
by a musician (one of the authors), wherever needed, 
to maintain consistency with the IDMT dataset.
The purpose of these two datasets is to shed light on the 
generalization capabilities of the methods, 
that is, with different instruments.

\subsection{Scoring Method}
For scoring, $\text{F}_{1}$ Score \cite{dixon2006onset, bock2013maximum} 
is computed.
The $\text{F}_{1}$ Score is the harmonic average of the 
Precision and Recall.
A value close to 1 is better and vice versa. 
Recall, also called as the true positive rate, 
in this scenario, is the fraction of the original onsets 
identified by the algorithm.  
Even if all time frames are identified to contain an onset, 
when in reality only a small percentage actually does,
recall would be maximum (100\%), since all positives (onsets) 
were accounted for. 
It does not take into account the number of
falsely identified positives. 
Hence, another parameter is needed to counter it --- Precision, 
also called positive predictive value. 
It is the fraction of the identified positives (onsets) 
that are actually true.

Discussions with respect to defining an 
objective location of an onset --- 
whether it should be defined as the point 
that denotes the beginning
of the transient, or its peak ---  
can be found in \cite{leveau2004methodology}
and \cite{collins2005comparison}.
In the current work, the point at the beginning 
of the transient is considered as the onset location, 
as this is the most commonly agreed upon definition 
\cite{bello2005tutorial, dixon2006onset}, 
and also because it makes most sense in a 
music information retrieval or an 
automatic music transcription framework.

An appropriate value of the error threshold
must also be set.
It is defined as the acceptable error 
in the accuracy of the detected onset 
with respect to the annotated or reference.
Error thresholds can affect the results drastically by
adding value to the accuracy and precision offered by an algorithm.
The most popular error threshold value in literature for the task of 
onset detection is $\pm$50ms \cite{bello2005tutorial} 
\cite{dixon2006onset} \cite{zhou2008music}.
This is to account for the fact that most of the datasets 
are hand-labelled \cite{bello2005tutorial}. 
The MIREX onset detection task specifications also 
specify an error threshold of 50ms \cite{mirexonset}. 
The current work assumes the same.
It can be argued that the threshold should be 
selected with respect to the task at hand,
considering parameters such as, the length of the average note that is produced by the instrument 
or the length of just the attack, 
and the general difficulty in accurately pinpointing 
the location of the onset. 
Some other good points of view can be found in \cite{leveau2004methodology}.

\subsection{Experimental Setup}
\label{experimental_setup}

All the algorithms that are evaluated in the current work
are briefed in Table \ref{tab:proposed_algorithms}. 
Each algorithm is encoded in column 1 
to provide details of each case. 
The first digit denotes the algorithm used to derive 
the OSS and corresponds to column 2 in the table,
the second digit corresponds to column 3, and so on. 
Henceforth, these algorithm numbers are used in 
all the tables that follow.

\begin{table}[h]
\begin{center}
\caption{Summary of the algorithms mentioned in the current work}
\vspace{0.25cm}
\begin{tabular}{ccccc}
\hline
Alg. No & OSS & HWR & CGD & Peak Picking \\ \hline 
1001            & Complex Domain Function                                                           & No                                                                         & No                                                                                       & PP1                                                             \\
1012            & Complex Domain Function                                                           & No                                                                         & Yes                                                                                      & VPD                                                             \\ 
2101            & Spectral Flux                                                            & Yes                                                                        & No                                                                                       & PP1                                                             \\
2112            & Spectral Flux                                                            & Yes                                                                        & Yes                                                                                      & VPD                                                             \\ 
3100            & SuperFlux                                                                & Yes                                                                        & No                                                                                       & PP2                                                             \\ 
4012            & Spectral Average                                                                  & No                                                                         & Yes                                                                                      & VPD                                                             \\ 
5000            & CNN-based Onset Detection                                                                  & No                                                                         & No                                                                                      & PP2                                                             \\ \hline
\end{tabular}
\label{tab:proposed_algorithms}
\end{center}
\end{table}

\begin{table}[]
\begin{center}
\caption{Evaluation of the chirp group delay algorithm as a smoothening  function, with respect to the IDMT dataset, Guitarset and Musicnet, in terms of $\text{F}_{1}$ Score, Precision and Recall}
\vspace{0.25cm}

\begin{tabular}{llcccc}
\hline
Dataset  & Algorithm & Alg No. & F1     & Recall & Precision \\ \hline
\multirow{4}{*}{IDMT} & CDF \cite{duxbury2003complex} & 1001    & 0.6118 & 0.6383 & 0.6329    \\
                              & CDF + CGD + VPD & 1012    & 0.8255 & 0.8003 & 0.8973    \\
                              & Spectral Flux \cite{masri1996computer} & 2101    & 0.8113 & 0.7622 & 0.9305    \\
                              & Spectral Flux + CGD + VPD & 2112    & 0.8327 & 0.8097 & 0.9078    \\ \hline 
\multirow{4}{*}{Guitarset}    & CDF \cite{duxbury2003complex} & 1001    & 0.7605 & 0.7584 & 0.7776    \\
                              & CDF + CGD + VPD & 1012    & 0.8579 & 0.8246 & 0.9058    \\
                              & Spectral Flux \cite{masri1996computer} & 2101    & 0.7685 & 0.6806 & 0.9298    \\
                              & Spectral Flux + CGD + VPD                             & 2112    & 0.8396 & 0.8524 & 0.8400    \\ \hline

\multirow{4}{*}{Musicnet}     & CDF \cite{duxbury2003complex} & 1001    & 0.4450 & 0.5393 & 0.4094    \\
                              & CDF + CGD + VPD & 1012    & 0.7604 & 0.8197 & 0.7401    \\
                              & Spectral Flux \cite{masri1996computer} & 2101    & 0.6214 & 0.5879 & 0.7010    \\
                              & Spectral Flux + CGD + VPD & 2112    & 0.7516 & 0.7278 & 0.8224 \\ \hline   
\end{tabular}

\label{tab:results_exp1}
\end{center}
\end{table}

Apart from the proposed valley-peak distance-based 
peak picking algorithm (VPD), 
two existing peak picking algorithms have been
used for comparison. 
The first one is explained in \cite{dixon2006onset}, 
and henceforth known as PP1. 
The other one used with superflux and CNN-based onset detection
is given in \cite{bock2012evaluating}, and henceforth known as PP2.
Both these algorithms compute a few threshold parameters 
at each time frame based on past, present, and future values, 
which are then compared to the current value of the OSS 
to make a decision if the time frame has a peak. 
These parameters, when plotted, are in some sense a 
lazy variant of the OSS itself by the nature of how 
they are defined mathematically with respect to the OSS ---
varying slowly, negating spurious smaller peaks while 
accommodating peaks with higher strengths to take precedence.
The peak picking algorithm used in each experiment is 
specified in Table \ref{tab:proposed_algorithms}.
They are paired based on existing experiments in 
\cite{dixon2006onset}, \cite{bock2012evaluating} and
\cite{schluter2014improved}.

To evaluate the proposed algorithms,
for OSS estimation, OSS processing and picking peaks,
explained in Section \ref{proposed_algorithms},
three streams of experiments are proposed.
\begin{itemize}
    \item First, in order to validate the capability of the 
    proposed chirp group delay algorithm as a smoothening function, 
    the OSS produced by other onset detection algorithms are processed
    by the CGD algorithm (Section \ref{chirp_group_delay_algorithm})
    and the peaks are picked by the VPD algorithm (Section \ref{peak_picking}).
    The two OSS considered are, 
    complex domain function and spectral flux.
    The results are compared with the complex domain function
    and spectral flux in their original form --- 
    1001 vs 1012, and 2101 vs 2112.
    For this stream, all three datasets are used.
    \item Secondly, the proposed algorithm which comprises
    the simpler short-time spectral average function 
    in unison with CGD and VPD algorithms, is validated
    as a viable alternative to the state-of-the-art.
    This algorithm is compared with two state-of-the-art algorithms,
    Superflux \cite{bock2013maximum} and 
    CNN-based onset detection (CNN-OD)\cite{schluter2014improved}.
    All three datasets are used in this stream too.
    \item Thirdly, once these two proposals are established, 
    all seven algorithms are compared with each other
    in terms of how they handle ornamentation. 
    For this, only the IDMT dataset is used.
    This third stream is split into two experiments based 
    on different abstractions of the IDMT dataset.
    One experiment is designed to compare the onset detection
    performance of all seven algorithms with respect to 
    the licks and scale-like pieces.
    The other experiment is designed to compare the 
    onset detection performance of all seven algorithms with 
    respect to the different ornamentation present in the licks.
\end{itemize}

For each of the seven methods in Table \ref{tab:proposed_algorithms}, 
the experimental parameters are tuned in such a way that 
the best overall score (as in Table \ref{tab:results_exp1} and 
\ref{tab:results_exp2})
is obtained across the whole dataset for that method.
Once this is done, the same parameter settings are used 
for all subsequent experiments on data subsets 
(Table \ref{tab:results_broadclasses} and \ref{tab:results_licks}).
The following experimental parameters are tuned: 
(i) the threshold parameter(s) of the peak picking algorithms, 
(ii) the radius of the chirp group delay estimation (1.001 to 1.020), and
(iii) the frame size (20 or 40 ms) and frame rate (5 or 10 ms) 
of the short-time Fourier transform.
For peak picking methods PP1 and PP2, 
the threshold parameter(s) are tuned as suggested 
in \cite{dixon2006onset} and \cite{bock2013maximum} respectively.
For VPD, the scaling factor $\mu$, 
that determines the threshold (eq. \ref{eq:threshold}),
is varied from 0.75 to 1.
All the signals are processed as monoaural signals, with a 
sampling rate of 44.1 kHz.

\begin{table}
\begin{center}
\caption{Comparison of the proposed onset detection algorithm with 
the state-of-the-art --- SuperFlux and CNN-based onset detection, with respect to the IDMT dataset, Guitarset and Musicnet, in terms of $\text{F}_{1}$ Score, Precision and Recall}
\vspace{0.25cm}

\begin{tabular}{llcccc}
\hline
Dataset  & \multicolumn{1}{c}{Algorithm} & Alg No. & F1     & Recall & Precision \\ \hline 
\multirow{3}{*}{IDMT}      & Superflux \cite{bock2013maximum} & 3100    & 0.8839 & 0.8578 & 0.9580    \\
                           & STSA + CGD + VPD & 4012    & 0.8890 & 0.8676 & 0.9555    \\
                           & CNN-OD & 5000    & 0.8741 & 0.8694 & 0.9104    \\ \hline
\multirow{3}{*}{Guitarset} & Superflux \cite{bock2013maximum} & 3100    & 0.9157 & 0.9035 & 0.9359    \\
                           & STSA + CGD + VPD & 4012    & 0.9117 & 0.8776 & 0.9543    \\
                           & CNN-OD & 5000    & 0.9295 & 0.9030 & 0.9605    \\ \hline
\multirow{3}{*}{Musicnet}  & Superflux \cite{bock2013maximum} & 3100    & 0.8550 & 0.8196 & 0.9206    \\
                           & STSA + CGD + VPD & 4012    & 0.8557 & 0.8416 & 0.8967    \\
                           & CNN-OD & 5000    & 0.9087 & 0.8825 & 0.9516   \\ \hline
\end{tabular}

\label{tab:results_exp2}
\end{center}
\end{table}

\subsection{Experimental Results}
\label{eval:exp_results}
\subsubsection{Evaluation of the chirp group delay algorithm as a smoothening function}
\label{eval:smoothening}
These experiments show the abilities of chirp group delay 
algorithm working as a smoothening function in unison with 
popular onset detection algorithms. 
Experiment pairs \textit{1001-1012}, and \textit{2101-2112},
in Table \ref{tab:proposed_algorithms} explain these experiments.
All three datasets, IDMT, Guitarset, and Musicnet, are analysed. 
Here, the entire IDMT dataset is used for evaluation, whereas 
Sections \ref{eval:broad_classes} and \ref{eval:licks}, 
use subsets of the IDMT Dataset.

The results of these experiments are shown in 
Table \ref{tab:results_exp1}. 
It can be inferred from the results that: 
\begin{itemize}
    \item when chirp group delay
    is used to smoothen the OSS generated by the 
    complex domain and the spectral flux functions, 
    there is consistent improvement in performance,
    across both algorithms and all datasets.
    \item when chirp group delay is used to smoothen the OSS generated 
    by the complex domain function, there is a considerable improvement of 
    35\% (0.6118 vs 0.8255), 13\% (0.7605 vs 0.8579), and 71\% 
    (0.4450 vs 0.7604), in the scores for the three datasets, respectively.
    \item when it is applied to the OSS generated by the spectral
    flux function, there is an improvement of 2.5\% (0.8113 vs 0.8327), 
    9\% (0.7685 vs 0.8396), 21\% (0.6214 vs 0.7516),
    in the scores for the three datasets, respectively. 
\end{itemize}
 
The comparatively lower $\text{F}_{1}$ score for Musicnet across all 
four methods can be attributed to the fact that the 
piano recordings contain various note lengths --- 
from `whole notes' that span around one second, to `sixteenth notes'
that span hundredths of that.
But it can be noticed that the improvements offered by the
chirp group delay-based smoothening algorithm is more in 
the case of Musicnet --- 71\% and 21\% for complex domain 
and spectral flux functions respectively.

\begin{figure*}[ht]
  \centerline{\includegraphics[width=0.75\textwidth]{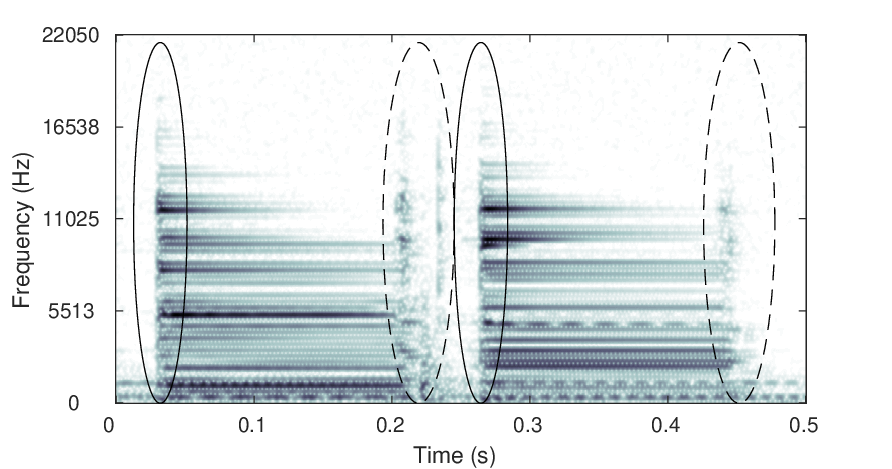}}
  \caption{Figure shows the spectrogram of two notes 
  to illustrate the occurrence of pseudo-onsets that 
  are caused by accidentally touching a vibrating
  guitar string, as the finger or plectrum is 
  moved away from it. 
  Ellipses with a solid line denote actual onsets, 
  while those with a dashed line, denote pseudo-onsets}
  \label{fig:false_positives}
\end{figure*}

The comparatively similar, but still mildly distinct, 
performances of IDMT and Guitarset must be pointed out.
Although both instruments (electric guitar in IDMT,
and steel-string acoustic guitar in Guitarset) have their similarities, 
steel-string acoustic guitars tend to have stronger onsets,
and it is tough to produce ornamentation in them.
Hence there is slightly better performance in all four cases, 
except for spectral flux.
The reason for a dip in the performance of spectral flux 
for Guitarset can be attributed to the fact that guitarists 
sometimes tend to touch a vibrating string, with the plectrum, 
as they move away from it.
This can produce narrow sharp peaks in the time-frequency
representation which look very similar to those of onsets.
Such pseudo-onsets were found in the Guitarset and 
they are illustrated in Fig. \ref{fig:false_positives}.
Spectral difference-based algorithms can be 
sensitive to this phenomenon.
This is a common scenario in guitar recordings, 
and many such instances were found in these recordings too.
It must be noticed that chirp group delay offers 
improvement in the $\text{F}_{1}$ scores in this case too.

\subsubsection{Comparison of the proposed onset detection algorithm with 
the state-of-the-art --- SuperFlux and CNN-based onset detection}
\label{eval:superflux}
When chirp group delay-based smoothening is applied to the OSS 
generated by STSA function, and peaks picked by VPD,
the performance is on par with the state-of-the-art. 
The results are shown in Table \ref{tab:results_exp2}.
The dataset remains the same as in the previous case.
The $\text{F}_{1}$ score obtained by superflux, 
CNN-based onset detection and the proposed method,
for the IDMT Dataset is 0.8839, 0.8741 and 0.8890 respectively.
While for Guitarset, they are 0.9157, 0.9295 and 0.9117, and
for Musicnet, 0.8550, 0.9087 and 0.8557, respectively.
It also performs better than all four algorithms 
discussed in the previous section.

\begin{table}
\begin{center}
\caption{Average computation time required to execute the OSS 
and Peak Picking algorithms for 3100 (superflux), 
4012 (proposed) and 5000 (CNN-based onset detection) in milliseconds}  
\vspace{0.25cm}
\begin{tabular}{cccc}
\hline
Algorithm & Alg. No.   & OSS (ms) & Peak Picking (ms) \\ \hline
Superflux \cite{bock2013maximum} & 3100              & 9.0          & 0.4                \\ 
STSA + CGD + VPD & 4012              & 3.0          & 0.2                \\
CNN-OD \cite{schluter2014improved} & 5000              & 1124.0          & 0.2                \\ \hline 
\end{tabular}
\label{tab:results_computation}
\end{center}
\end{table}

Table \ref{tab:results_computation} shows the average time 
(in milliseconds) it takes to execute the OSS and 
the Peak Picking algorithms for each file in the dataset. 
All algorithms were implemented in Python and 
tested on a computer running on an Intel Core i5-460M processor, 
with no other applications running in the background.
Although the onset estimation performance of 
these three algorithms is on par, 
it can be seen that the proposed algorithm 
is more efficient than the state-of-the-art.
The difference in computation times can be explained
based on the steps of the algorithm. 
Both the proposed and the superflux  algorithms 
require the estimation of STFT. 
But, the max filtering operation in superflux 
is performed on a two dimensional STFT magnitude spectrum
before being reduced to the one dimensional OSS signal. 
In contrast, the chirp group delay processing 
is performed on the reduced one dimensional OSS signal.
On the other hand, 
even though CNN is a versatile supervised classification scheme,
both the proposed and superflux are 
much more efficient for the current task.

\begin{table}
\begin{center}
\caption{Experimental results over two broad subsets (licks and scales) of the IDMT Dataset, in terms of $\text{F}_{1}$ Measure, Precision and Recall}
\vspace{0.25cm}

\begin{tabular}{clcccc}
\hline
        & Algorithm & Alg No. & F1 & Recall & Precision \\ \hline
\multirow{7}{*}{\rotatebox[origin=c]{90}{Licks}}  & CDF \cite{duxbury2003complex} & 1001                        & 0.4615                 & 0.4889                     & 0.5013                        \\
                        & CDF + CGD + VPD                                & 1012                        & 0.8229                 & 0.7889                     & 0.9181                        \\
                        & Spectral Flux \cite{masri1996computer}            & 2101                        & 0.7674                 & 0.7043                     & 0.9220                        \\
                        & Spectral Flux + CGD + VPD                                          & 2112                        & 0.8223                 & 0.7886                     & 0.9232                        \\
                        & Superflux \cite{bock2013maximum}                  & 3100                        & 0.8691                 & 0.8266                     & 0.9728                        \\
                        & STSA + CGD + VPD                                                   & 4012                        & 0.8633                 & 0.8303                     & 0.9550                        \\
                        & CNN-OD                                                             & 5000                        & 0.8713                 & 0.8481                     & 0.9297                        \\ \hline
\multirow{7}{*}{\rotatebox[origin=c]{90}{Scales}} & CDF \cite{duxbury2003complex} & 1001                        & 0.8146                 & 0.8400                     & 0.8106                        \\
                        & CDF + CGD + VPD                                & 1012                        & 0.8340                 & 0.8361                     & 0.8321                        \\
                        & Spectral Flux \cite{masri1996computer}            & 2101                        & 0.9492                 & 0.9444                     & 0.9570                        \\
                        & Spectral Flux + CGD + VPD                                          & 2112                        & 0.8651                 & 0.8761                     & 0.8593                        \\
                        & Superflux \cite{bock2013maximum}                  & 3100                        & 0.9305                 & 0.9558                     & 0.9116                        \\
                        & STSA + CGD + VPD                                                   & 4012                        & 0.9697                 & 0.9849                     & 0.9572                        \\
                        & CNN-OD                                                             & 5000                        & 0.8827                 & 0.9366                     & 0.8495   \\ \hline                    
\end{tabular}
\label{tab:results_broadclasses}
\end{center}
\end{table}

A valid conclusion mentioned in \cite{dixon2006onset} 
must be reiterated here: 
``differences in F-measure between the best algorithms 
are not significant, implying that the choice of 
algorithm could be based on other factors such
as simplicity of programming, speed of execution 
and accuracy of correct onsets". 
This conclusion goes well with our argument that 
the chirp group delay offers performance on par 
with the state-of-the-art, whilst maintaining lower 
computational complexity.

\subsubsection{Evaluation of all algorithms on 
the two broad subsets in the IDMT dataset}
\label{eval:broad_classes}
The two broad subsets of the IDMT dataset are licks and scales, 
as explained in Section \ref{corpus}. 
The performance of all seven algorithms with respect to these two 
broad subsets is given in Table \ref{tab:results_broadclasses}.

It can be seen that in the case of licks, which contain 
all the ornamentation mentioned earlier, 
the vanilla complex domain and spectral flux functions
do not perform well due to reasons already discussed. 
But chirp group delay offers an improvement of 
approximately 78\% (0.4615 vs 0.8229) and 
7\% (0.7674 vs 0.8223) respectively.
On the other hand, it can be seen that in the case of 
scale-like pieces, the case is different --- 
the baseline algorithms perform well.
The improvement offered by chirp group delay for the 
complex domain OSS is comparatively minimal. 
In the case of spectral flux, though, there is no improvement; 
in fact, vanilla spectral flux performs better. 
This is because a very clear distinction between notes
is already present in these recordings. 
In any case, it must also be remembered that this subset 
does not generalize real world recordings at all.

Algorithms, 3100, 4012 and 5000 perform on par with each other, 
and better than all the other algorithms mentioned 
in the previous section,
when detecting onset locations for the licks. 

\begin{table}
\centering
\caption{Experimental results with respect to different 
playing styles in the IDMT Dataset, in terms of $\text{F}_{1}$ Measure. The styles are encoded in the following manner, B - Bend, S - Slide, 
H - Harmonics,  V - Vibrato, D - Dead and, N - Normal.}
\vspace{0.25cm}
\begin{tabular}{lccccccccc}
\hline
Style & Files - Onsets & 1001 & 1012 & 2101 & 2112 & 3100 & 4012 & 5000           \\ \hline
BSH                                  & 9 - 87                   & 0.6448        & 0.9368        & 0.8828        & 0.9227        & 0.9825        & 0.9566        & 0.9385 \\
BVDN                                 & 9 - 81                   & 0.5337        & 0.9519        & 0.7642        & 0.9494        & 0.9877        & 0.9431        & 0.9533 \\
HBV                                  & 9 - 72                   & 0.5378        & 0.9804        & 0.9208        & 0.9869        & 0.9935        & 0.9746        & 0.9360 \\
N                                    & 99 - 2196                 & 0.4455        & 0.7907        & 0.7272        & 0.7744        & 0.8240        & 0.8204        & 0.8479 \\
NVSBHD                               & 9 - 603                  & 0.4954        & 0.6008        & 0.7420        & 0.8066        & 0.9810        & 0.9206        & 0.9770  \\
SBDN                                 & 9 - 117                  & 0.5081        & 0.9041        & 0.8599        & 0.9218        & 0.9867        & 0.9877        & 0.9914 \\
VSDN                                 & 9 - 117                  & 0.5362        & 0.9911        & 0.9057        & 0.9079        & 0.9702        & 0.9822        & 0.9712 \\
VSH                                  & 9 - 15                   & 0.5313        & 0.9410        & 0.9053        & 0.9669        & 0.9935        & 0.9746        & 0.9617 \\
N\_Lage                              & 27 - 519                  & 0.4662        & 0.8379        & 0.8496        & 0.8394        & 0.8935        & 0.9043        & 0.8937 \\
N\_Lage2                             & 9 - 96                   & 0.4191        & 0.7955        & 0.8500        & 0.8600        & 0.9464        & 0.9314        & 0.9442 \\ \hline
\end{tabular}

\label{tab:results_licks}
\end{table}

\subsubsection{Evaluation of all algorithms on the various 
lick styles in the IDMT dataset}
\label{eval:licks}

Performance of all seven algorithms, with respect to
the ten different playing styles is also computed 
and provided in Table \ref{tab:results_licks}. 
These playing styles are subsets of the broad subset ‘Licks’, 
in the IDMT dataset.
A similar trend exists here too. 
Chirp group delay offers considerable improvement to the 
complex domain function and a fair amount of improvement 
in the case of spectral flux.
When comparing algorithms 3100, 4012 and 5000, 
we notice that they offer in general, similar performances.

Two things should be pointed out here. 
First, although a particular data subset shown in 
Table \ref{tab:results_licks} may be tagged
to contain a certain ornamentation, not all, 
but only one or two notes in these files contain it. 
Second, the scores can fluctuate quite easily based 
on the number of files, or the
number of onsets in each subset (which varies from 15 to 2196). 
The smaller the number of onsets, the greater the difference 
that individual errors will make in the score. 
Such weighting needs to be
given when considering the scores in each category. 
Table \ref{tab:results_licks} provides the number of files 
and the number of onsets in those files for each playing style 
to gain this perspective.

\section{Conclusion}
\label{conclusion}

A computationally efficient algorithm to process and 
refine the OSS derived from a time-frequency representation 
using chirp group delay is proposed. 
Such processing yields favorable characteristics in the OSS that
suppress the chance of obtaining false positives and false negatives.
It also makes it easier to pick the peaks from the OSS function, 
since the number of peaks that the peak picking algorithm 
needs to eliminate as spurious is less with smoothening.
The smoothening capability of chirp group delay 
is evaluated by processing other major OSS functions,
namely complex domain and spectral flux functions, 
before peak-picking, and investigating the results. 
In both these cases when the OSS of these functions 
are smoothened using chirp group delay, 
considerable improvement in performance is recorded.

An onset detection algorithm that involves computing the OSS using 
short-time spectral average function, and then 
smoothening it with chirp group delay, 
and peaks picked by valley-peak distance 
based peak picking algorithm, is proposed. 
The proposed algorithm performs on par with the state-of-the-art, 
superflux and CNN-based onset detection, 
and is computationally more efficient.
It offers an efficiency improvement of around 300\% 
with respect to superflux, 
which can be very useful since an onset detection
algorithm is used repeatedly as a low-level task. 
It is evident from these experiments that reducing the 
load on the peak picking algorithm by better shaping 
the OSS can provide better results.

The performance of these algorithms when applied to different 
playing styles and different instruments is also investigated. 
It can be noted that for each category, the proposed algorithms
perform on par with the state-of-the-art
and offer considerable improvement over 
vanilla complex domain and spectral flux functions.

\paragraph{Data Availability Statement:}
The datasets used for analysis in the current study are available from the following links.
\begin{itemize}
    \item IDMT-SMT-Guitar ---  \url{https://www.idmt.fraunhofer.de/en/publications/datasets/guitar.html}
    \item Guitarset --- \url{https://guitarset.weebly.com/}
    \item Musicnet --- \url{https://zenodo.org/record/5120004#.YhYWoXVBzMU}
\end{itemize}

\end{document}